\documentstyle[11pt]{article}
\newcommand{\be}{\begin{equation}}
\newcommand{\ee}{\end{equation}}
\newcommand{\ba}{\begin{eqnarray}}
\newcommand{\ea}{\end{eqnarray}}
\newcommand{\bann}{\begin{eqnarray*}}
\newcommand{\eann}{\end{eqnarray*}}

\begin{document}
\hbadness=10000
\setcounter{page}{1}
\title{
\vspace{-3.0cm}
\hspace{-1.0cm}
\hspace*{\fill}
{\normalsize DMR-THEP-95-1/W} \\*[3.0ex]
{Production of a chaotic squeezed state from a ``pion liquid" and
overbunching of identical pion correlations}
}

\author{I.V. Andreev$^1$ \thanks{E.mail: andreev@lpi.ac.ru} and
R.M. Weiner$^2$ \thanks{E.mail: weiner@mailer.uni-marburg.de}}

\date{$^1$ P.N. Lebedev Physical Institute, Moscow, Russia\\
$^2$  Physics Department, University of Marburg, Marburg, F.R. Germany}

\maketitle

\begin{abstract}
It is shown that a one to one correspondence between quantum fields in two
different ``phases" as might be realized for pions produced from a ``hadron
liquid" leads to squeezed states. The single and double inclusive cross
sections for
at chaotic superposition of such states are calculated.
The correlation of identical pions is
overbunched in comparison with canonical Bose-Einstein correlations.
\end{abstract}

Recent experimental data \cite{adler} on identical pion correlations in
nucleon-antinucleon annihilation in rest indicate that the intercept of the
second order correlation function $C_2(p,p)$ reaches values much larger than
the
``canonical" maximum value of 2 observed in other reactions. This effect which
we shall call ``overbunching" was interpreted in ref. \cite{song} in
terms of
resonance production. In this interpretation the space-time information which
identical particle interferometry provides through the analogy with the Hanbury
Brown-Twiss (HBT) effect in optics, i.e. Bose-Einstein correlations (BEC) is
lost. While the resonance production is certainly a factor to be considered for
annihilation processes, in the present paper an alternative mechanism for
the overbunching
effect will be proposed in terms of squeezed states, which maintains the link
with the HBT effect and which may have other implications in particle
physics.\footnote{Note that as shown in \cite{bolz} resonance production in
itself does
not necessarily spoil the conventional interpretation of identical particles
correlations in terms of BEC. This is the case only for particular models like
that considered in \cite{song}.}
At the same time the derivation of the correlation function for squeezed states
given below is a new result also in quantum optics.

We consider pion production via the decay of a stronly interacting pionic
system. We suggest that, due to fluctuations, a blob of pionic matter created
in particle collision, undergoes a sudden breakup into free pions. In other
words, the pionic system, having its specific ground state and elementary
pionic
excitations (not coinciding exactly with the usual vacuum and free particles)
converts rapidly into free pions. In this case the single and higher order
inclusive cross section and the many-particle correlation functions will depend
on the spectrum of excitations in the pionic system. The importance of the form
of the spectrum of pionic excitations for multiparticle production was stressed
some time ago in ref. \cite{shury} where arguments were given for a softening
of
the spectrum of pions in hot hadronic matter, that is, for a given momentum
$\vec{p}$, the effective pion energy in the medium was asserted to be less than
the free pion energy.

We shall argue below that the above physical picture results in the production
of quantum squeezed states. These states represent a $U(1,1)$ group
extension of the coherent states, allowing for incertitudes in momentum
(coordinate) below those obtained for coherent states (cf. e.g. \cite{walls}).
It is known that squeezing may lead to an enhancement of particle
correlations, but so far the second order correlation function $C_2(p_1,p_2)$
is known only for pure squeezed states.
The creation of this kind of states was conjectured recently in
ref. \cite{kogan} in the framework of a disoriented chiral pion
condensate. Unlike ref. \cite{kogan}, we shall consider
the more general case that the pion system may be (partially) chaotic (hot)
and thus an averaging over pion states has to be performed.
A chaotic distribution is equivalent to a Gaussian form of the
density matrix in the $P$ representation (cf. below) which in its turn follows
in the assumption that the number of independent sources is large. The coherent
and chaotic distributions are the two most interesting extremes in particle
physics and quantum optics (cf. \cite{v} where the chaotic distribution of
squeezed states, i.e. the ``thermal" squeezed states were introduced).

We consider the transition from a pionic ``liquid" to a free pion field in the
spirit of local parton-hadron duality, i.e. we conjecture a close
correspondence
between particles (fields) in the two ``phases".
 At the moment of this transition we postulate
the following relations between the generalized coordinate $Q$
and the generalized momentum $P$ of the field:
\ba
Q &=& \frac{1}{\sqrt{2E_b}} (b^+ + b) = \frac{1}{\sqrt{2E_a}} (a^+ + a)
\nonumber\\
P &=& i \sqrt{\frac{E_b}{2}} (b^+ - b) = i \sqrt{\frac{E_a}{2}} (a^+ - a)
\label{eq:qp}
\ea
$a^+,a$ are the free field creation and annihilation operators and $b^+,b$ the
corresponding operators in the ``liquid". Eq. (\ref{eq:qp}) holds for each mode
$p$. Then we get immediately a connection between the $a$ and $b$ operators,
\ba
a &=& b \; cosh \; r \; + \; b^+ \; sinh \; r \; , \nonumber\\
a^+ &=& b \; sinh \; r \; + \; b^+ \; cosh \; r \;
\label{eq:aa}
\ea
with
\be
r = r (\vec p) = \frac{1}{2} \, log \, (E_a/E_b) \, .
\label{eq:rr}
\ee

The transformation (2) is just the squeezing transformation \cite{walls} with
a momentum dependent squeezing parameter $r(\vec p)$ given by eq. (\ref{eq:rr})
and the coherent eigenstate $|\beta>_b$ of the $b$-operator is the squeezed
state $|\alpha,r>_a$ of the $a$-operator:
\be
|\beta>_b = |\alpha,r>_a
\label{eq:bb}
\ee
where $\alpha$ and $\beta$ are related by the same transformation (2) as the
$a$
and $b$ operators.

In general the system may not be in a pure coherent or squeezed state and then
a
statistical averaging has to be performed both with respect to the coherent as
well as for the squeezed states. In practice it is easier to express the
$a,a^+$-operators through the $b,b^+$-operators according to eq. (\ref{eq:aa})
and then perform the averaging over the coherent states $|\beta>_b$.
Considering
charged
identical pions (complex valued field) we shall use the Glauber-Sudarshan
representation \cite{glauber} of the density matrix, and write the average
value
of an operator $\hat{O}$ as
\be
<\hat O (a,a^+) > = \prod \limits_{\vec p} \int d^2 \beta_k P \{ \beta(\vec p)
\} <\beta|\hat{O} \, (a(b,b^+),a^+(b,b^+))|\beta>_b
\label{eq:oo}
\ee
and assume a Gaussian form for the weight function $P\{\beta(\vec p)\}$. Due to
linearity of the squeezing transformation (2) this form will hold also for the
$a,a^+$-operators.

A blob of decaying pionic matter (particle source) will be characterized by a
primordial correlator determined by the
number density $n(p)$ and by a function $f(\vec{x})$ describing its
geometrical form, see
refs. \cite{aw,apw}. To make contact with the previous results of \cite{aw,apw}
we note that the radius of the source $R$ enters the function $f$ and the
correlation length $L$ appears in $n(p)$. For simplicity we shall not consider
the time dependence here and take the form function $f(\vec x)$ to be dependent
only on the space coordinates.

The direct substitution of the transformation (2) into eq. (\ref{eq:oo}) leads
to undefined (divergent) expressions of the form $\delta(0)$
when one tries to perform the normal
ordering of $b,b^+$-operators (the last is necessary to use the coherent state
representation of eq. (\ref{eq:oo})). To avoid this situation we introduce new
creation and annihilation operators which are non-zero only inside the volume
of
the particle source,
\be
\tilde a(\vec x) = a (\vec x) f (\vec x) \quad , \quad \tilde{a}^+(\vec x) =
a^+ (\vec x) f (\vec x) \; ,
\label{eq:aa2}
\ee
or, for Fourier transformed quantities,
\be
\tilde a (\vec p) = \int \frac{d^3k}{(2\pi)^3} a(\vec k) f(\vec k - \vec p)
\quad , \quad \tilde{a}^+ (\vec p) = \int \frac{d^3k}{(2\pi)^3} a^+(\vec k)
f(\vec p - \vec k)
\label{eq:aa3}
\ee
with standard commutation relations
\be
\left[ a(\vec{p}_1),a^+ (\vec{p}_2) \right] = (2\pi)^3 \cdot
\delta^3(\vec{p}_1-\vec{p}_2) \, .
\label{eq:aa4}
\ee

Then the equal momentum commutators of the modified operators are finite. For
example:
\be
[\tilde a (\vec p),\tilde{a}^+ (\vec p)] = \int \frac{d^3k}{(2\pi)^3}
f(\vec p - \vec k)f(\vec k - \vec p) = \int d^3x |f|^2(\vec x) = V_{eff}
\label{eq:aa5}
\ee
being equal to an effective volume $V_{eff}$ of the particle source.
While this finite size is quite natural in particle physics, it is not
so in optics where the system is usually macroscopic.

With the smoothed operators $\tilde a(\vec p),\tilde{a}^+(\vec p)$ substituted
into eq. (\ref{eq:oo}) the form of the source is already taken into account and
the remaining statistical averaging may be performed in the same way as for
an infinite medium. We have checked that the above procedure reproduces
correctly previous results of refs. \cite{aw,apw} on correlation functions (at
least for the case when the time dependence is omitted).

Now the evaluation of the averaged matrix elements is straightforward.
Substituting eqs. (\ref{eq:aa3}) and (\ref{eq:aa}) into eq. (\ref{eq:oo}) and
performing the Gaussian averaging over coherent states $|\beta>_b$ we get the
single-particle inclusive density in the form:
\ba
\rho_1(\vec p)  &=& \frac{(2\pi)^3}{\sigma} \frac{d\sigma^{in}}{d^3p} =
<\tilde{a}^+(\vec p) \tilde{a}(\vec p)> \nonumber\\
&=& \int \frac{d^3k}{(2\pi)^3} \left[ n_b(\vec k) cosh \, 2r(\vec k) + sinh^2 r
(\vec k) \right] f(\vec p - \vec k) f(\vec k - \vec p)
\label{eq:rr2}
\ea
where the function $f$ describes the effect of finite size of the particle
source and $n_b(\vec k)$
given by the equation
\be
<\beta^*(\vec{k})\beta(\vec{k}')> = (2\pi)^3 \delta^3 (\vec{k}-\vec{k}')
n_b(\vec{k})
\label{eq:new}
\ee
represents the density of pionic ``quasiparticles"
($b$-quanta) (in particular, for a thermal source the
function $n_b(\vec k)$ is the usual Planck distribution function).

The squeezed state effect is reflected in eq. (\ref{eq:rr2}) in the factor
$cosh2r$ multiplying the primary pionic density
$n_b(\vec k)$ and in the term $sinh^2r$ representing the ground state
contribution. That is the final state pions are produced even if the pions in
the pionic source are absent (zero temperature), just due to the decay of the
squeezed vacuum state. According to eq. (\ref{eq:rr2}), the single particle
density may be strongly enhanced in the presence of squeezed states if the
squeezing parameter $r(\vec p)$ is large enough.

We consider now the two-particle inclusive density
\be
\rho_2(\vec{p}_1,\vec{p}_2) = \frac{(2\pi)^6}{\sigma} \cdot \frac{d \sigma}
{d^3p_1 \cdot d^3p_2} = <\tilde{a}^+(\vec{p}_1) \tilde{a}^+(\vec{p}_2)
\tilde{a}(\vec{p}_1) \tilde{a}(\vec{p}_2)>
\label{eq:rr3}
\ee
in the presence of squeezed states. With the finite size cut off the
two-particle density is calculated in
the same way as the single-particle density. Using Gaussian averaging
one gets the simple expression
\be
\rho_2(\vec{p}_1,\vec{p}_2) = <\tilde{a}^+(\vec{p}_1) \tilde{a}(\vec{p}_1)>
<\tilde{a}^+(\vec{p}_2) \tilde{a}(\vec{p}_2)> +
|<\tilde{a}^+(\vec{p}_1) \tilde{a}(\vec{p}_2)>|^2 + |<\tilde{a}(\vec{p}_1)
\tilde{a}(\vec{p}_2)>|^2
\label{eq:rr4}
\ee
with
\ba
<\tilde{a}^+(\vec{p}_1) \tilde{a}(\vec{p}_2)> &=&
\int \frac{d^3k}{(2\pi)^3} \left[ n_b(\vec k) cosh \, 2r(\vec k) + sinh^2 r
(\vec k) \right] f(\vec{p}_1 - \vec k) f(\vec k - \vec{p}_2), \nonumber\\
<\tilde{a}(\vec{p}_1) \tilde{a}(\vec{p}_2)> &=&
\int \frac{d^3k}{(2\pi)^3} \left[ n_b(\vec k) + \frac{1}{2} \right] sinh \, 2r
(\vec k) f(\vec{k}_1 - \vec{p}_1) f(\vec k - \vec{p}_2) \, ,
\label{eq:aa6}
\ea

The first term in the right hand side of eq. (\ref{eq:rr4}) is the product of
single-particle densities $\rho_1(\vec{p}_1) \rho_1(\vec{p}_2)$, the second
term
is the exchange contribution characteristic for Bose-Einstein correlations
modified by the squeezing factor $r$ (for $r = 0$ it coincides with the
usual BEC).
The third term arises only in the presence of squeezed states (it
vanishes for $r = 0$). This last contribution differs from the ``surprising"
terms in the two-particle correlation function discussed in refs.
\cite{apw,apw2}, which are absent in the case of charged identical pions under
consideration and which have another dependence on momenta
$\vec{p}_1,\vec{p}_2$, being maximal at $\vec{p}_1 + \vec{p}_2 = 0$, and not at
$\vec{p}_1 - \vec{p}_2 = 0$ as is the case for all terms in eq. (\ref{eq:rr4}).

As one can see from eqs. (\ref{eq:rr2}), (\ref{eq:rr4}),(\ref{eq:aa6}) the
second order correlation function
\be
C_2(\vec{p}_1,\vec{p}_2) = \rho_2(\vec{p}_1,\vec{p}_2)/\rho_1(\vec{p}_1)\rho_1
(\vec{p}_2) \, ,
\label{eq:rr5}
\ee
is enhanced due to the presence
of the third term in the right hand side of eq. (\ref{eq:rr4}) and in general
the value of the ratio (\ref{eq:rr5}) is arbitrarily large. In particular, for
$n_b(\vec k) = 0$ (that is for cold pionic matter when particle production is
the result of the squeezed vacuum decay) and for small values of the
squeezing parameter
$r(\vec{k})$, one may have $C_2 >> 1$ (however, under these conditions
the single particle inclusive cross-section (\ref{eq:rr2}) is very
small). For $r(\vec{k}) \sim 1$ and $\vec{p}_1 \cong \vec{p}_2$ the ratio
(\ref{eq:rr5}) is close to three.




In conclusion we see that the investigation of Bose-Einstein correlations and
in
particular of the overbunching effect may shed light on the
the role of sqeezing in pion production and on the momentum dependence
of the squeezing parameter which in its turn provides information about the
spectrum of pions in pionic matter.

{\bf Acknowledgements}\\
This work was supported in part by the Deutsche Forschungsgemeinschaft and the
Russian Fund for Fundamental Research, Project No. 93-02-3815.
We are indebted to M. Pl\"umer for instructive discussions and for his
participation in the early stages of this work. Useful discussions with Y.
Sinyukov are also gratefully acknowledged.


\begin{thebibliography}{99}
\bibitem{adler} B. Adler et al., CPLEAR Coll., Z.Phys. {\bf C63} (1994) 541.
\bibitem{song} H.Q. Song et al., Z.f.Phys. {\bf A342} (1992) 439.
\bibitem{bolz} J. Bolz et al., Phys.Rev. {\bf C47} (1993) 3860.
\bibitem {shury} E.V. Shuryak, Phys.Rev. {\bf D42} (1990) 1764.
\bibitem{walls} D.F. Walls, G.J. Milburn, ``Quantum optics", Springer Verlag
Berlin, Heidelberg, 1994.
\bibitem{kogan} I.J. Kogan, JETP Lett. {\bf 59} (1994) 307.
\bibitem{v} A. Vourdas, Phys.Rev. {\bf A34} (1986) 1986; A. Vourdas and R.M.
Weiner, Phys.Rev. {\bf A36} (1987) 5866; A. Vourdas and R.M. Weiner, Phys.Rev.
{\bf D38} (1988) 2209.
\bibitem{glauber} R.J. Glauber, Phys.Rev. {\bf 131} (1963) 2766. E.C.G.
Sudarshan, Phys.Rev.Lett. {\bf 10} (1963) 277.
\bibitem{aw} I.V. Andreev, R.M. Weiner, Phys.Lett. {\bf 235} (1991) 416.
\bibitem{apw} I.V. Andreev, M. Pl\"umer, and R.M. Weiner, Int.Journ.Mod.Phys.
{\bf A8} (1993) 4577.
\bibitem{apw2} I.V. Andreev, M. Pl\"umer, and R.M. Weiner, Phys.Rev.Lett.
{\bf 67} (1991) 3475.
\end{thebibliography}
\end{document}